\documentclass[12pt]{article}
\usepackage{amsmath,bm}
\usepackage{amssymb}
\usepackage{graphicx}
\usepackage{amsfonts}         
\usepackage{slashed}
\usepackage{yfonts} 
\usepackage{verbatim}
\usepackage{wrapfig}

\renewcommand{\baselinestretch}{1.2}
\let\non\nonumber
\renewcommand{\theequation}{\arabic{section}.\arabic{equation}}

\global\newcount\itemno \global\itemno=0

\def\itemaut#1{\global\advance\itemno by1\noindent\item{\the\itemno.}#1}

\newif{\ifeq}           
\eqtrue                 

\newcommand{\be}{\begin{equation}}
\newcommand{\ee}{\end{equation}}
\newcommand{\bes}{\begin{equation*}}
\newcommand{\ees}{\end{equation*}}
\newcommand{\bea}{\begin{eqnarray}}
\newcommand{\eea}{\end{eqnarray}}
\newcommand{\bean}{\begin{eqnarray*}}
\newcommand{\eean}{\end{eqnarray*}}

\def\({\left(}
\def\){\right)}
\def\[{\left[}
\def\]{\right]}

\def\frac#1#2{{#1 \over #2}}

\newcommand{\rt}{{\sqrt 2}}

\renewcommand{\a}{\alpha}
\renewcommand{\b}{\beta}
\renewcommand{\d}{\delta}
\newcommand{\g}{\gamma}

\newcommand{\s}{\sigma}

\renewcommand{\l}{\lambda}

\newcommand{\e}{\epsilon}
\newcommand{\eps}{\epsilon}

\renewcommand{\t}{\theta}

\newcommand{\G}{\Gamma}

\newcommand{\U}{\Upsilon}
\newcommand{\LL}{\Lambda}

\newcommand{\CA}{{\cal A}}

\newcommand{\CJ}{{\cal J}}

\newcommand{\CL}{{\cal L}}

\newcommand{\CN}{{\cal N}}
\newcommand{\CO}{{\cal O}}

\newcommand{\CV}{{\cal V}}

\newcommand{\IC}{{\mathbb C}}

\newcommand{\IP}{{\mathbb P}}

\newcommand{\IZ}{{\mathbb Z}}

\newcommand{\Tr}{{\rm Tr \,}}
\newcommand{\tr}{{\rm tr \,}}

\def\Re{{\rm Re\hskip0.1em}}
\def\Im{{\rm Im\hskip0.1em}}

\def\ie{{\it i.e.}}

\def\etal{{\it et. al.}}

\newcommand{\lsim}{\,\raise.3ex\hbox{$<$\kern-.75em\lower1ex\hbox{$\sim$}}\,}
\newcommand{\gsim}{\,\raise.3ex\hbox{$>$\kern-.75em\lower1ex\hbox{$\sim$}}\,}

\def\p{\partial}

\def\susy{supersymmetry}
\def\susic{supersymmetric}

\def\Ka{K\"{a}hler}

\def\nK{non-K\"{a}hler}

\def\ZT{$(0,2)$}

\def\II{\relax{I\kern-.10em I}}

\newcommand{\ttheta}{\theta^+}
\newcommand{\tthetabar}{\bar{\theta}^+}
\newcommand{\Ree}{\mbox{Re} \,}
\newcommand{\Imm}{\mbox{Im} \,}
\newcommand{\open}{\left( \begin{array}{cc} }
\newcommand{\opens}{\left( \begin{array}{cccc} }
\newcommand{\close}{\end{array}\right)}

\begin{document}
\begin{titlepage}
\begin{flushright}
MIT-CTP 4012
\end{flushright}

\vskip 1in
\begin{center}

{\Large Heterotic Flux Vacua from Hybrid Linear Models}

\vskip 0.6in Allan Adams and David Guarrera
\vskip 0.4in {\it Center for Theoretical Physics \\ Massachusetts Institute of Technology \\ Cambridge, MA  02139 USA}

\end{center}

\vskip 0.8in

\begin{abstract}
\noindent
We construct hybrid linear models in which the chiral anomaly of a gauged linear sigma model is canceled by the classical anomaly of a gauged WZW model.  Semi-classically, this corresponds to fibering the WZW model over the na\"ive target space of the sigma model.  When the gauge group is abelian, we recover known \nK\ compactifications; non-abelian models describe novel quasi-geometric flux vacua of the heterotic string.
\end{abstract}

\vfill 
\begin{flushleft} 
\today 
\end{flushleft}
\vfil

\end{titlepage}
\newpage

\renewcommand{\baselinestretch}{1.1}  
\renewcommand{\arraystretch}{1.5}

\section{Introduction}

One of the missing links in our understanding of string compactifications is a microscopic description of spacetimes stabilized by fluxes.  One reason this is hard is that most of the familiar flux vacua are solutions of Type II with Ramond-Ramond fluxes turned on.  While beautiful in many ways, such solutions are largely intractable using traditional worldsheet techniques\footnote{Considerable progress has been made in quantizing RR backgrounds by Berkovits and collaborators using the pure spinor and hybrid formalisms.  However, a generally applicable and computationally effective formalism analogous to the GLSM remains elusive.  For now, RR vacua remain challenging.}.  To avoid these complications, we might try to quantize pure NS-NS vacua of Type II.  Unfortunately, turning on $H$-flux generically generates tree-level tadpoles which can only be cancelled by decompactifying or adding orientifolds and other RR objects, so that doesn't solve the problem.  

The difficulties of quantizing RR fluxes can be avoided by working in a heterotic duality frame, where a tree-level $H$-flux can be balanced against a 1-loop anomaly via the Green-Schwarz mechanism,
$$
dH = \a' \(tr ~\!R\wedge R ~-Tr~\! F\wedge F\).
$$
We thus have a fighting chance of constructing a worldsheet CFT for heterotic flux vacua.  Unfortunately, this equation is abrasively non-linear, making the construction of concrete non-Calabi-Yau solutions exceedingly challenging.

Considerable progress was made on this problem with the identification of a special class of non-trivial solutions in \cite{Dasgupta:1999ss,Fu:2005sm, Fu:2006vj,Becker:2006et}.  These solutions all take the form of $T^{2}$-fibrations over a base $K3$, with $H$-flux along the fibration balancing against the curvature of the bundle so as to satisfy the Bianchi identity above.  While these vacua have $c_{3}\(\CV\)=0$, and thus have zero generations at the semi-classical level, they provide interesting toy models of non-trivial heterotic flux compactifications.

These vacua were subsequently realized and generalized on the worldsheet via chiral gauged linear sigma models in \cite{Adams:2006kb} in which a 1-loop gauge anomaly cancels the classical gauge anomaly of a set of dynamical axions.
This mechanism is the pullback to the worldsheet of the spacetime Green-Schwarz effect.  The fact that all such models have zero generations follows\footnote{We thank J. Lapan for discussions on this point.}, in the worldsheet description, from the existence of a pair of free right-moving fermions (the superpartners of the axions, which are coordinates on the $T^{2}$ fiber) whose zero modes ensure that all spacetime fermions come in non-chiral pairs.  For these and other reasons, it would be interesting to generalize these models beyond the original example of $T^{2}$-fibrations over \Ka\ manifolds.
	
The goal of this note is to construct one such generalization.  To introduce $H$-flux, we again pull back spacetime Green-Schwarz anomaly cancellation to a GLSM for the worldsheet CFT.  This time, however, we will not require the worldsheet anomaly to be abelian.  So long as we are careful to keep all possible anomalies cancelled, making the gauge group non-abelian boils down to replacing the $T^{2}$ fiber with some non-abelian group, $G$ (or, more generally, some coset $G/H$), a subgroup of which is identified with the gauge group of the GLSM.  More precisely, rather than starting with an anomalous abelian GLSM and canceling the anomaly by coupling the theory to a scalar axion in a gauge-non-invariant fashion, we now start with an anomalous non-abelian gauge theory and cancel the anomaly by coupling to a classically-anomalous gauged WZW model.  By suitable choice of coset, we can ensure that there are no free right-moving fermions to force the spacetime spectrum to be chiral -- these vacua do not, in general, have generation number zero.  The result is a hybrid WZW gauged linear sigma model providing a worldsheet description of a large class of new quasi-geometric heterotic flux vacua which reduces to the original $T^{2}$ fibration in the abelian case.

Notably, something very similar was done in a pair of beautiful 
papers by Johnson \etal\ \cite{Johnson:1994kv,Berglund:1995dv}, who built novel $(0,2)$ ``minimal models'' by adding $(0,2)$-singlet left-moving fermions to gauged WZW models so as to cancel the one-loop anomaly generated by the fermions against the classical anomaly of the WZW model.  One of the mysteries of those models was where, on the moduli space of string vacua, they arose; one lesson of this line of work is that they arise on the moduli space of \nK\ flux-vacua of the heterotic string.  A similar strategy was also used in a recent paper by Distler and Sharpe \cite{Distler:2007av}, who built WZW-fibered non-linear sigma models over Calabi-Yau 3-folds to realize $E_{8}$ bundles over topological CYs which could not be otherwise realized via free fermions.

It would be interesting to compute the spectra and chiral rings of these models. As discussed in \cite{Adams2009}, this is naturally done by flowing to a Landau-Ginzberg point of the GLSM, e.g. at $r \to$ -$\infty$, where the partition function of the full theory reduces to an orbifold of the product of the LG partition function with the WZW partition function. Viewed as a symmetry in either the LG or the WZW theory individually, the orbifold group is anomalous; when the partition functions are taken together, the anomaly cancels. This is the exact conformal field theory avatar of gauge anomaly cancellation in the UV GLSM/WZW hybrid. 

This paper is organized as follows.  In Section 2 we quickly review $\CN=2$ gauged WZW models.  In Section 3 we couple such WZW models to anomalous gauged linear sigma models to cancel the gauge anomaly of the GLSM.  In Section 4 we identify the necessary non-anomalous $U(1)_{L}$ and $U(1)_{R}$ symmetries needed for a computation of the spectrum.  In Section 5 we discuss how some of our models may be obtained by bosonization and fermionization.  We then introduce several explicit examples in Section 6 and conclude in Section~7.

\newcommand{\Gg}{{\textswab{g}}}

\section{Gauged WZW Models} \label{WZWintro}

We briefly review the gauged WZW models described in \cite{Johnson:1994kv} and \cite{Berglund:1995dv}.  
A WZW model with $(0,1)$ supersymmetry contains $G$-valued scalar bosons, $g(x) \in G$, together with right handed Majorana-Weyl superpartners, $\psi_+$, valued in ${\textswab{g}} = T_{G}$, the Lie algebra of $G$.  To gauge the WZW model we introduce two vector fields, $A^{L}$ and $A^{R}$, gauging $H_{L,R} \subset G$, where $H_{L}$ and $H_{R}$ are generated by left and right multiplication, 
\bea
g &\rightarrow& h_L^{}  ~\! g ~\!  h_R^{-1} \non \\
\psi_+& \rightarrow& h_R^{}  ~\! \psi_+  ~\! h_R^{-1} \non \\
A^{L} &\rightarrow& h_L^{}  ~\! dh_L^{-1} + h_L^{}  ~\! A^{L}  ~\! h_L^{-1}  \non\\
A^{R} &\rightarrow& h_R^{} ~\!  d h_R^{-1} + h_R^{}  ~\! A^{R}  ~\! h_R^{-1}. \non
\eea
The action of the gauged model is then,
\bea\label{WZW}
S = &-&\!\!{k \over 4\pi} \int_{\Sigma} \tr\!\[g^{-1} \partial_+ g g^{-1} \partial_- g\] -i\tr\!\[ \psi_+ D_- \psi_+\]  \non\\
&-& \!\!{ik\over 2\pi} \int_{\Sigma} \tr\!\[g^{-1} \partial_+ g A^{R}_- - A^{L}_+ \partial_- g g^{-1} + iA^{R}_- g^{-1} A^{L}_+ g + {i \over 2}(A^{L}_+ A^{L}_-+A^{R}_+A^{R}_-)\]    \non \\
&-&\!\!\!{ik \over 12 \pi} \int_V   \tr\!\[(g^{-1} \partial_i g)(g^{-1} \partial_j g)(g^{-1} \partial_k g)\] \e^{ijk}
\eea
where $V$ is a volume bounded by the worldsheet and $D_- \psi_+= \p_- \psi_+ -i [A^{R}_-, \psi_+]$ is the covariant derivative of our right-moving fermions, which take values in the algebra of the coset $G/(H_R \times H_L)$, \ie,
$$
\psi_+ \in Lie(G)-Lie(H_R)-Lie(H_L).
$$
The model actually has $(0,2)$ supersymmetry if the coset satisfies the following conditions:
\begin{itemize}
\item{$T_{\mathbb{C}}$, the Lie algebra of $G/(H_R \times H_L)$,  has the decomposition $T_{\mathbb{C}}=T_+ \oplus T_-$ of conjugate representations. This is the statement that $G/(H_R \times H_L)$ has a local complex structure.}
\item{$[T_+, T_+] \subset T_+$. and $[T_-, T_-] \subset T_-$. This is the statement that the Nijenhuis tensor vanishes and the complex structure is integrable.}
\item{$\tr\!(ab)=0$ if $a,b \in T_+$ or $a,b \in T_-$. This is the statement that there exists a hermitian $(1,1)$ form on $G/(H_R \times H_L)$.}
\end{itemize}
Under these conditions, the model is invariant under the $(0,2)$ SUSY transformations
\begin{eqnarray} \label{WZWSusy}
\delta g&=&i \epsilon_1 g \psi_{R-} + i \epsilon_2 g \psi_{+} \nonumber \\
\delta \psi_{+}&=&\epsilon_1 \Pi_+ (g^{-1} D_+ g - i \psi_{+} \psi_{-}-i \psi_{-} \psi_{+} ) +i \epsilon_2 \psi_{+} \psi_{+} \nonumber \\
\delta \psi_{-}&=&\epsilon_2 \Pi_- (g^{-1} D_+ g - i \psi_{+} \psi_{-}-i \psi_{-} \psi_{+} ) +i \epsilon_1 \psi_{-} \psi_{-} \nonumber \\
\delta A^{L}_{\pm}&=&0 \nonumber \\
\delta A^{R}_{\pm}&=&0 
\end{eqnarray}
where $\Pi_\pm$ is the projection to $T_{\pm}$ and $D_{\pm} g \equiv \partial_{\pm} g -i A^{L}_{\pm} g+i g A^{R}_{\pm} $.  For unitary groups with $g^{-1}=g^\dagger$, consistency of the SUSY transformations requires that $\epsilon_2=-\bar{\epsilon}_1$. 

Finally, and crucially for our later purposes, this action is in fact classically anomalous for a general gauging: under a gauge transformation with left/right gauge parameters $\alpha_L$, $\alpha_R$, the action shifts by,
\be \label{classanomaly}
\delta S=\frac{k}{4 \pi} \(\tr\! [\alpha_R \, F^{R'}_{+-}]-\tr\! [\alpha_L \, F^{L'}_{+-}]\) ,
\ee
with $F^{R}$ the field strength for $A^{R}$ and $F^{L}$ the field strength for $A^{L}$ (the primes indicate that only the $dA$ terms in $F$ appear, \ie\ only the ``consistent" anomaly contributes).

\subsection{An Example} \label{WZWexample}
Let's examine a simple example, the $SU(2)/U(1)$ WZW model, where the $U(1)$ generated by $\sigma_3/2$ has been gauged on the right (\ie\ with $A^{R}$).  Since $SU(2)/U(1) \sim \IP^{1}$ is complex and hermetian (in fact, \Ka), this model should admit a \ZT\ supersymmetric extension.  As it stands, however, the Lagrangian is not gauge-invariant.  To cancel this classical anomaly, we introduce left handed fermions charged under the $A^{R}$ gauge symmetry; these chiral fermions generate a quantum anomaly which cancels the classical anomaly of the gauged WZW model\footnote{This specific theory has been used in the construction of worldsheet theories that describe four dimensional heterotic solutions of a black hole of magnetic charge $Q$, \cite{Giddings:1993wn}.}.  The anomaly cancellation condition is then
$$ 
{k\over2} +1-Q^{2}=0, 
$$
where the $k/2$ is the coefficient of the classical anomaly of the WZW model (the $1/2$ is from the normalization of the generators of $SU(2)$), the $+1$ comes from the Weyl fermion in the WZW model with gauge charge +1, and the $-Q^2$ is the contribution from the left handed fermion with charge $Q$. 

Significantly, since left-handed fermions are singlets (on-shell) under right-moving \ZT\ supersymmetry, adding them does not spoil \ZT\ \susy.  These models are known as $(0,2)$ minimal models \cite{Berglund:1995dv}\ and have central charge 
\be \label{stringent}
c= \frac{3k}{k+2} \,.
\ee
As an application, we can use these minimal models to build realistic heterotic compactifications with $c=9$. Condition (\ref{stringent}) is a very restrictive condition on $k$ -- so restrictive, in fact, that the only way to build a vacuum with the correct central charge is to take the tensor product of four theories with $k=6$ ($Q=2$)  \cite{Berglund:1995dv}. Of course, one may generate more possibilities by taking the tensor product of this model with $(2,2)$ WZW models -- these are the so called ``doped" models  of \cite{Berglund:1995dv}.

\section{Constructing the Hybrids}
As reviewed in the last section, we can cancel the quantum anomaly generated by a set of chiral fermions by coupling in a gauged WZW model, with total anomaly cancellation imposing a single condition relating the charges of the fermions to the level of the WZW model.  In this section we will study a natural generalization of this mechanism in which we replace the chiral fermions by a gauged linear sigma model whose fermion content is anomalous.  In this more intricate case, vanishing of the net anomaly will again reduce to a set of conditions relating the charges of the matter fields in the gauge theory to the level of the WZW model.  Studied semiclassically, the net effect will be to fiber the WZW model non-trivially over the classical target space of the sigma model.

Two points need to be kept firmly in focus.  First, neither the gauge theory nor the WZW model is independently invariant under the symmetry we would like to gauge -- the gauge theory suffers from a quantum anomaly and the WZW model is classically anomalous.  It is only the combination of the two which realizes this symmetry exactly and allows us to gauge.  Second, both the gauge theory and the gauged WZW are independently supersymmetric, despite the anomalies.  This is obscured when working in WZ gauge, where the SUSY algebra closes only up to a gauge transformation and thus does not close in the presence of a gauge anomaly.  However, this is a failure of WZ gauge, not of supersymmetry, and is in any case of no concern so long as we focus on the non-anomalous combination of gauge theory and gauged WZW model.

\subsection{The Gauge Theory}

Let's start with a very brief review of the non-abelian $(0,2)$ gauged linear sigma model that will occupy us in what follows.  For a detailed review of $(0,2)$ GLSMs, see \cite{Distler:1995mi,Witten:1993yc}.  We start by introducing $N$ chiral multiplets $\Phi_{i=1\ldots N}$, 
$$
\Phi_i =   \phi_i + \sqrt{2} \theta^+ \psi_{+i}-i \theta^+ \bar{\theta}^+ \partial_+ \phi_i 
$$
transforming in representations $R_i$ of a symmetry group $G$, together with $M$ chiral Fermi multiplets $\Gamma_{a=1\ldots M}$,
$$
\Gamma_a = \gamma_{-a}  - \sqrt{2} \theta^+ F_a -i \theta^+ \bar{\theta}^+ \partial_+ \gamma_{-a} \non \\
$$
transforming in representations $R_a$ of $G$.  Since $F_{a}$ is auxiliary, the left-moving fermions are on-shell supersymmetry singlets.  This is the full matter sector of the GLSM.

We introduce dynamics by gauging $G$ with a \ZT\ vector multiplet, $V_{\pm}$.  In components, 
\bea \label{general}
V_-&=& A_-  - 2 i \theta^+ \bar{\lambda}_-- 2 i \bar{\theta}^+ \lambda_- + 2 \theta^+ \bar{\theta}^+ D \non \\
V_+&=& C + i \ttheta \gamma_+ + i \tthetabar \bar{\gamma}_+ + \ttheta \tthetabar A_+ \,.
\eea
The canonical field strength supermultiplet
\bea
\Upsilon_{-} &=& [e^{V_+} \bar{\mathcal{D}}_+ e^{-V_+}, \nabla_{-}] \non \\
&=&(-2 \lambda_- -i D_- \g_+) + 2 i \ttheta (D + {i\over 2} F_{+-}+ ...) + 2 i \ttheta \tthetabar (D_+ \lambda_-+ ...) \non
\eea
transforms in the adjoint of $G$, where $(...)$ denotes terms that will shortly be set to zero by a choice of gauge and $\nabla_{-}=\p_-+iV_{-}$ is the left-moving gauge-covariant superderivative.

Under a supergauge transformation with adjoint-valued chiral gauge parameter
\be
B=b + \sqrt{2} \ttheta \b_{+} -i \ttheta \tthetabar \partial_+ b,
\ee
the matter fields transform according to their representations while $V_{\pm}$ transform as,
\bea
V_{+} &\to& V_+ + i (B-\bar{B}) -i [V_+, B+\bar{B}] + \ldots \non\\
V_{-} &\to& V_- + i\p_{-} (B-\bar{B}) + i [V_+, B+\bar{B}] + \ldots \non
\eea
In components, the variation of $V_{+}$ takes the form,
\bea
C &\to& C-2i \mbox{Im} \, b -i [C, 2 \Ree b] + \ldots  \non \\
\gamma_+ &\to& \gamma_+ + \sqrt{2} \b_{+} -i [\gamma_+, 2 \Ree b] - \sqrt{2} [C, \b_{+}] + \ldots  \\
A_+ &\to& A_+ +2 \partial_+ \Ree b -i [A_+, 2 \Ree b] + \sqrt{2} [\gamma_+, \bar{\b}_{+}] + \sqrt{2} [\bar{\gamma}_+, \b_{+}] -i [C, \partial_+ 2 \Imm b] +\ldots ,\non
\eea
where the $(\ldots)$ terms involve higher order commutators involving $\Imm b$. 


As in four dimensions, we can use our super-gauge invariance  to fix the non-dynamical components of $V_{+}$ to zero\footnote{Specifically, taking $b-\bar{b} = iC$ sets $C\to0$, while taking $\b_{+}=-\g_{+}$ subsequently sets $\g_{+}\to0$.}, leaving $V_{+}$ in the form,
$$ 
V_+=\ttheta \tthetabar A_+.
$$
This so-called Wess Zumino (WZ) gauge is particularly intuitive, since it makes manifest that the only propagating degrees of freedom in the vector multiplet are the gauge boson $A_{\pm}$ and the gaugino, $\lambda_-$ (which resides in $V_{-}$).  Notably, since any gauge transformation with $\b_{+}=0$ and $\Im b=0$ preserves the WZ condition, WZ gauge-fixing preserves a residual unfixed gauge symmetry, $A_+ \rightarrow A_+ + D_+ a$, where $a=2\Re b$. These are just the usual gauge transformations associated with any gauge theory. 

Sadly, the benefits of WZ gauge come at a cost.  By fixing some of the components of vector superfield $V$ to zero, we have destroyed manifest supersymmetry.  Explicitly, under a SUSY transformation with SUSY parameter $\e$, the vector $V_{+}$ transforms out of WZ gauge,
\be \label{firstpsi}
V_{+} ~ \to ~ i \theta^{+}\bar{\e} A_+ -i \bar{\theta}^{+}\e A_+ + \ttheta \tthetabar A_+ \,.
\ee
We can return to WZ gauge by making a further gauge transformation with gauge parameter,
\be \label{WZLambda}
B^{WZ} = {-i\over\rt}\( \theta^{+}\bar{\e} A_+ - \bar{\theta}^{+}\e A_+\) .
\ee
It is easy to check that this returns us to WZ gauge.  The theory is thus only supersymmetric up to a gauge transformation in WZ gauge.  As long as our theory is gauge-invariant, this is a nitpicking detail.  Happily, the canonical Lagrangian for our theory,
\bea
\CL &= &\int \! d^{2}\t \,\Tr\!\[\,
\frac{1}{8 e^2}\bar{\U}\U - \frac{i}{2} 
\sum_{i} e^{V_{+}}\bar{\Phi_{i}}\nabla_{-}e^{V_{+}}\Phi_{i}-\frac{1}{2}
\sum_{a} e^{V_{+}}\bar{\G}_{a}e^{V_{+}}\G_{a} 
\,\] \non \\
&+& \frac{1}{4} \int d\t^{+} \Tr\!\[ \,  t\U \, \] + h.c. 
\eea
is indeed classically gauge invariant, so we appear to be safe.

\subsection{Gauge Anomalies in the GLSM}

Since our theory contains fermions transforming in chiral representations of the gauge group, it runs the risk of a chiral anomaly which spoils gauge-invariance.  Without gauge invariance, negative norm states no longer decouple and unitarity is lost. We must proceed with caution.

In two dimensions the anomaly comes from a diangle diagram with one current insertion and one external gauge boson.  The derivation (see for example \cite{AlvarezGaume:1983ig}) of this anomaly proceeds as in four dimensions.  In a $U(1)$ gauge theory with $N$ right-handed fermions of charge $Q_i$ and $M$ left-handed fermions of charge $q_a$, the chiral anomaly of the gauge current, $J_G^\mu$, under variation with gauge parameter $\alpha$, is
\be
\partial_\mu J_G^\mu=\frac{\CA}{2 \pi}  \alpha F_{+-},
\ee
where the anomaly coefficient is given by $\CA=\sum_{i}Q_i^2-\sum_{a}q_a^2$. For a non-abelian theory with semi-simple gauge group, this generalizes to\footnote{The rather annoying factor of 2 between the abelian and non-abelian anomalies derives from different conventional normalizations of the generators.}
\be \label{GLSMAnomaly}
\partial_\mu J_G^\mu=\frac{\mathcal{A}}{4 \pi} \Tr [\alpha F_{+-}]
\ee
with $\mathcal{A}$ again determined by the matter fields and their representations. (Nonabelian anomalies will be discussed in more detail in section \ref{examples}.) 

In 2d theories enjoying $(2,2)$ \susy, this anomaly always vanishes, since every right handed fermion lives in a supermultiplet with a left-handed partner, so both transform in the same gauge representation. Said differently, $(2,2)$ supersymmetry only allows for matter that is in a non-chiral representation of the gauge group. 

In the \ZT\ theories of interest to us, left- and right-chiral fermions live in different representations (chiral and fermi, respectively) of \ZT\ \susy, and may thus transform in different representation of the gauge group.  We should thus expect a \ZT\ supersymmetric extension of this anomaly in our theories.  For semi-simple Lie groups, the resultant super anomaly is
\be
\frac{\mathcal{A}}{4 \pi} \int d \ttheta  \Tr\!\[ B \Upsilon\] +\mbox{h.c.} \,,
\ee
where $B$ is the gauge parameter and $\Upsilon$ the gauge field strength supermultiplet.  

Without further modification, only models whose chiral anomalies vanish make sense. 
Nevertheless, let us for the moment soldier on and consider $(0,2)$ models with non-vanishing chiral anomaly. 
This leads to an important subtlety with WZ gauge, where SUSY is only respected up to a gauge transformation: if the anomalous theory is not invariant under gauge transformations, the theory in WZ gauge would not appear to be supersymmetric.  Explicitly, suppose we perform a supersymmetry transformation with parameter $\e$, then apply the WZ-restoring gauge transformation (\ref{WZLambda}).  The resultant shift in the action is found by evaluating the anomaly on this gauge variation,
$$
{\CA\over 4 \pi} \int d \ttheta  \Tr\!\[ B^{WZ} \Upsilon\] .
$$
Since this is non-vanishing for general $\Upsilon$, supersymmetry appears broken in WZ gauge.  Of course, this is purely an artifact of fixing WZ gauge -- if we do not fix WZ gauge, the action is explicitly SUSY-invariant without any additional gauge transformation.  Nonetheless, ensuring that this ``WZ anomaly'' eventually cancels will be a useful check of the gauge-invariance of what follows.

\subsection{Adding the WZW Theory} \label{merging}

In section \ref{WZWintro}, we used the classical anomaly of a gauged WZW model to cancel the quantum chiral anomaly of a charged Weyl fermion.  As we have just seen, the quantum anomaly of a general GLSM takes the same form as these earlier anomalies.  This suggests a simple way to construct new non-anomalous models by balancing the classical gauge anomaly of a WZW model against the chiral anomaly of a $(0,2)$ GLSM.  As we shall see, the total theory can indeed be made non-anomalous and well defined. 

At first glance, there are a number of choices to be made in coupling the GLSM to the WZW model.  Explicitly, the GLSM contains a single dynamical vector, $A$, transforming non-trivially under supersymmetry.  The WZW model, by contrast, boasts two non-dynamical vectors, $A^{L}$ and $A^{R}$, which transform trivially under supersymmetry.   If our goal is to play the quantum anomaly of one off the classical anomaly of the other, they must be coupled to the same vector.  We thus must identify $A$ with either $A^{L}$ or $A^{R}$.  In the WZW model, this means promoting one of $A^{L,R}$ to a dynamical vector transforming non-trivially under supersymmetry.  So: which do we pick?

Supersymmetry guides our choice.  In the GLSM in WZ gauge, $A_{+}$ transforms trivially under SUSY while $A_{-}$ transforms non-trivially, with $\delta A_{-} = 2i \e \l_{-}$ (cf eq (\ref{general})).  Meanwhile, the vector couplings in the WZW model take the form (cf eq (\ref{WZW})),
$$
-{ik\over 2\pi} \int_{\Sigma} \tr\!\[g^{-1} \partial_+ g A^{R}_- - A^{L}_+ \partial_- g g^{-1} + iA^{R}_- g^{-1} A^{L}_+ g + {i \over 2}(A^{L}_+ A^{L}_-+A^{R}_+A^{R}_-)\] 
$$
If we promote $A^{R}$ to a dynamical field with the same SUSY variations as $A$, the WZW action will pick up a term proportional to $g^{-1}\p_{+}g$ under SUSY variation.  To preserve SUSY, this must cancel against some other term in the action.  Unfortunately, no other vector coupling in the action has a $g$-dependent SUSY variation.  So $A^{R}$ is out.

By contrast, if we promote $A^{L}$ to a dynamical field, the variation of the WZW action, while non-zero due to the $A^{L}_{-}A^{L}_{+}$ term, is at least independent of $g$ and thus has a chance of being cancelled by something in the GLSM action.  Explicitly, the action varies by a term proportional to the SUSY parameter, $\e$, the right-chiral boson, $A_{+}$, and the level, $k$, of the WZW model.  The component form is easily worked out to be,
\be
\d\CL_{WZW}=-{ik\over 2 \pi}  \(\epsilon \, \tr\![A_+ \bar{\lambda}_- ] + \bar{\epsilon} \, \tr\![A_+ \lambda_-]\) .
\ee
Worryingly, this does not look like the SUSY variation of any term in the GLSM action, so we again look stuck.


At this point something beautiful happens.  Recall that our GLSM in WZ gauge is not in fact \susic, but picks up a non-trivial SUSY-variation due to the anomaly of the WZ-restoring gauge shift.  Evaluating this explicitly gives,
\be
\d \CL_{gauge}={\CA \over 4 \pi}  \, 2i \(\epsilon \, \Tr\![A_+ \bar{\lambda}_- ] + \bar{\epsilon} \, \Tr\![A_+ \lambda_-]\) .
\ee
Delightfully, the form of the resulting variation precisely matches the SUSY variation of the WZW model!  Requiring that the total variation of the action vanishes then imposes a single condition relating the level, $k$, to the anomaly, $\CA$, of the GLSM,
\be \label{cancellationone}
k \, \tr\![T^2] = \CA  \, \Tr\![T^2]  ,
\ee
where $tr$ denotes the trace in the WZW model and $Tr$ the trace in the gauge theory, which may involve a different normalization (an endless source of spurious factors of 2).

The beauty of this condition is that its satisfaction ensures not just supersymmetry but also cancellation of the total anomaly.  Under a left-gauge variation (\ref{classanomaly}) the WZW model picks up a classical variation,
\be
\d\CL_{WZW}=-{k \over 4 \pi}\, \tr\![\alpha \, F_{+-}'] ,
\ee
while the GLSM picks up the chiral anomaly (\ref{GLSMAnomaly}), so that the total anomaly is,
\bea
\d\CL &=& \d\CL_{gauge} + \d\CL_{WZW} \non\\
&=& {\CA \over 4 \pi} \Tr\![\alpha \, F_{+-}'] - {k \over 4 \pi} \, \tr\![\alpha \, F_{+-}'] .
\eea
Requiring that the total anomaly vanish thus imposes the same condition as that needed for manifest supersymmetry.

At this point, we have succesfully cancelled the anomaly of our gauge theory by coupling in a classically anomalous WZW model.  However, several points deserve further comment.   First, we have been cavalier about the role of WZ gauge in the above.  As noted, both the WZW model and the gauge theory are independently supersymmetric.  However, once we fix to WZ gauge in which SUSY is only a symmetry up to a gauge variation, neither model is manifestly supersymmetric due to the (classical, quantum) anomaly generated by the gauge transformation needed to restore WZ gauge.  Thus WZ-gauge SUSY invariance really is nothing other than a measure of anomaly cancellation.

Secondly, while we have discussed in some detail the fate of the vector $A_{L}$ gauging the left-action, we have not mentioned that of $A_{R}$ gauging the right-action.  In particular, even if we do not promote it to a dynamical vector (which we shan't, as this would upset both gauge-invariance and supersymmetry), so long as we take $H_{R}$ to be non-trivial, $A_{R}$ will still couple to an anomalous current.  The crucial observation here is that we can always include SUSY-singlet left-moving fermions to cancel this anomaly without altering any of the considerations above.

Finally, as we originally introduced it, our WZW model for the coset $G/(H_{R}\times H_{L})$ contained right handed fermions living in $Lie(G)-Lie(H_R)-Lie(H_L)$ and coupled to the non-dynamical $A_{L}$.  In our hybrid GLSM, however, $A_{L}$ is only non-dynamical in the deep IR where the gauge coupling runs strong; at finite energy, the vector field is dynamical and the bosonic field $g$ lives in $G/H_{R}$.
Correspondingly, $\psi_{+}$, the right-handed superpartners of $g$, begin life valued in $Lie(G)-Lie(H_R)$ in the UV, with the restriction to $Lie(G)-Lie(H_R)-Lie(H_L)$ in the IR coming from their coupling to the gauginos. The full theory is supersymmetric iff $G/H_R$ is a complex manifold with hermitian metric.

\section{$U(1)_R$ and $U(1)_L$ Symmetries}

Every \ZT\ superconformal theory contains a purely right-moving conserved $U(1)$ current, $\CJ_{R}$, whose OPE determines the central charge, 
$$
\CJ_{R}(z) \CJ_{R}(0) = {\hat{c}_R\over z^{2}} + \dots
$$
One of the virtues of the GLSM is that we can often identify a candidate conserved R-current in the UV which flows to purely-right-moving conserved current $\CJ_{R}$ in the IR.  By 't~Hooft anomaly matching and asymptotic freedom, we may thus (in principle) compute the central charge of the strongly-coupled  IR theory by computing weak-coupling OPEs in the UV.  A similar story obtains for the left-moving current, $\CJ_{L}$.

Our goal in this section is to identify conserved $U(1)_R$ and $U(1)_L$ currents in the UV which flow to purely right/left-moving conserved currents in the IR.  In canonical $(0,2)$ GLSMs, it suffices to assign $U(1)_R$ charges to the matter fields compatable with the superpotential such that the $R$-current $\CJ^{\mu}_R$ is non-anomalous, conserved, and orthogonal to all non-anomalous flavor currents.  In general, the resulting $\CJ^+_R$ contains terms that either flow away in the IR or whose divergence is  trivial in $Q_+$ cohomology, so that the on-shell current runs to the holomorphic conserved current of the IR superconformal algebra. 

For our gauged WZW+LSM hybrids, identifying the correct $R$-currents is a little more subtle.   Na\"ively, the thing to do is assign each field a general $R$-transformation law, compute the resulting current by varying the action according to this symmetry, and deduce what the $R$-transformations must be for the resulting current to transform as an $R$-current.  Without loss of generality, we can assign $g$ the charge $Q T^{(1)}$, $\psi_{+}$ the charge $Q_R T^{(2)}$ and $\psi_{-}$ the charge $-Q_R T^{(2)}$, where $T^{(1,2)}\in\Gg$ specify the embedding of our $U(1)$ in $\Gg$, and the $Q$'s are real numbers.  
We can then try to construct a conserved $R$-current by varying the action according to this symmetry.
However, due of the classical non-gauge invariance of the WZW action, the resulting current, $\CJ_{R}$, is in general neither conserved nor gauge invariant.  To identify a good $R$-current, we will need to modify this na\"ive current to preserve gauge invariance (in a manner very similar to \cite{Hori:2001ax}).  

\def\wtilde{\widetilde}

For example, let's take an abelian model with $G=U(1) \times U(1)$.  The right-moving fermions $\psi_{+i}$ in the chiral multiplets $\Phi_{i}$ carry $U(1)_R$ charge $q^{i}_{R}$, the left-moving fermions $\l_{-}^{a}$ in the fermi multiplets $\Gamma^{a}$ carry $U(1)_R$ charge $q^{a}_{R}$, the right-moving fermions $\psi_{+}$ in the WZW multiplet carry shift charge $+1$, and the bosons $\theta^{l=1,2}$ in the WZW model carry shift charge $q_l$.  The corresponding na\"ive currents are,
\bea
\CJ_R^+ & = & \frac{1}{2 e^2} \lambda_- \lambda_- +q^R_a \bar{\l}_{-a} \l_{-a} -\frac{k}{4 \pi} q_l \partial_- \theta_l      \nonumber \\
\CJ_R^- & = &  q^R_i \bar{\psi}_{+i} \psi_{+i} + \frac{k}{4 \pi} \bar{\psi}_{+} \psi_{+} - \frac{k}{4 \pi}  q_l \partial_+ \theta_l   + \frac{k q_l N_l}{2 \pi} A_+ .
\eea
As expected, these currents are not gauge invariant, nor is it clear that the divergence of $\CJ_R^+$ is $Q$-trivial. 
Happily, it is easy to identify their (classically) gauge invariant cousins as
\bea 
\wtilde{\CJ}_R^+ & =& \CJ_R^+ + \frac{k q_l N_l}{4 \pi} A_- \nonumber \\
\wtilde{\CJ}_R^- & =& \CJ_R^+ - \frac{k  q_l N_l}{4 \pi} A_+ \, .
\eea
The $q_l$ are then chosen such that, if $\CJ_G$ is the gauge current, the leading term in the $\CJ_G^+ \wtilde{\CJ}_R^+$ OPE is equal to that of the $\CJ_G^- \wtilde{\CJ}_R^-$ OPE (this is the same as requiring that $\wtilde{\CJ}_R$ is gauge invariant quantum mechanically). The leading coefficient of the  $\wtilde{\CJ}_R^- \wtilde{\CJ}_R^-$ OPE minus that of the  $\wtilde{\CJ}_R^+ \wtilde{\CJ}_R^+$ will give $\hat{c}$, one third of the central charge of the right moving scft.   

Since  $e^{2}$ runs strong in the IR, the contributions of the left handed gaugini to $\wtilde{\CJ}_R^+$ flow away in the IR, while the $U(1)_R$ charged fermi multiplets develop masses.  The divergence of the remaining part of $\wtilde{\CJ}_R^+$ is then,
\bea
\p_+ \wtilde{\CJ}_R^+ & = & \ldots - \frac{k q_l}{4 \pi} ( \partial_+ \partial_- \theta_l -N_l \partial_+ A_-) \nonumber \\
&=& \ldots + \frac{k q_l N_l}{4 \pi} F_{+-} \nonumber \\
&\propto& \ldots + \{Q_+, \lambda_-\}
\eea
where in the second line we have used the $\theta$ equation of motion, and in the third we have used the SUSY algebra (for portions of the moduli space with $D=0$).  Since this is $Q$ trivial, we thus expect $\wtilde{J}_R^{+}$ to flow away completely so that $\wtilde{J}_R^{-}$ is the holomorphic, right moving $R$-current in the deep IR.


For non-abelian $G$, we expect a similar story -- the $U(1)_R$ and $U(1)_L$ currents will be a sum of the individual GLSM and WZW currents (for a general WZW model, these will involve the Lie algebra fermions and the Kac-Moody currents), corrected by $A$ dependent terms to preserve gauge invariance.  

Before moving on, it is useful to emphasize the apparent latitude we have in building specific examples.  Recall that cancellation of the gauge, $U(1)_R$, $U(1)_L$, and mixed $U(1)_R$/$U(1)_L$ anomalies of a conventional \ZT\ GLSM, plus the requirement that the low energy central charge is an integer, ensures that the target space is Calabi Yau. When anomaly cancellation is ensured by fibering a WZW model over a gauge theory, the ability to assign various $U(1)$ charges to the fibers would seem to free us from the requirement that the base be CY.  Of course, in that case the FI parameter of the GLSM runs, and a detailed understanding of the IR CFT requires a more nuanced analysis than the brief discussion above.  In the remainder of this note we will focus on the simplest case, in which the base is a CY; it would be interesting to explore the fate of more general examples with Ricci-curved bases.

\section{An Alternate Construction: Bosonization}

WZW models were originally discovered \cite{Witten:1983ar}\ as an answer to the question: ``What is the bosonization of an equal number of left and right moving fermions?"   For example, $N$ right and left moving Majorana-Weyl fermions may be bozonized into a $k=1$, $O(N)$ WZW model. 

The fibred models discussed above also arise via a combination of bosonization and fermionization. In these models, however, the fermion spectrum is chiral, so we must consider the bosonization and fermionionization of chiral systems (see, for example, \cite{Garousi:1995mq}.)   While straightforward, the process is not pretty.  

For example, consider the \ZT\ cft given by the tensor product of a free $T^2$ sigma model (at free-fermion radius) and a non-anomalous abelian GLSM with target space $K3$.  The basic strategy is to fermionize the free left-chiral bosons in the $T^{2}$ multiplet (this gives a set of free left-moving Weyl fermions) while bosonizing a pair of gauged left-moving fermions in the GLSM (this gives a set of left-gauged chiral bosons which, together with the original free right-chiral bosons, form a left-gauged $U(1)\times U(1)$ WZW model at $k=2$ \cite{Witten:1983ar,Witten:1991mk}).  The resulting model is thus the original GLSM coupled to a left-gauged WZW model and free fermions -- ie, the $T^{2}$ is now fibred over the base, while the fermions are trivial lines.  By construction, the contribution to the quantum gauge anomaly of the original left-handed fermions is now generated by the classical anomaly of the gauged WZW model.  This is just the left-gauged WZW-fibred GLSM discussed above.

More generally, we can start with a simple \ZT\ GLSM with gauge group $G_{GLSM}$ and target space $X$ and tensor on a WZW model for $G_{WZW}$ (perhaps with additional right-gauging by some $H_{R}\subset G_{WZW}$).  Now bosonize some subset of charged left-moving fermions in the GLSM whose contribution to the anomaly lies in $G_{Anom}\subset G_{GLSM}\cap G_{WZW}$ and fermionize the left-chiral bosons of the WZW model such that the final left- and right-chiral bosons form a $G_{WZW}/(G_{Anom} \times H_{R})$ WZW model, with the dualized left-moving fermions uncharged under the vector of the GLSM.  The quantum anomaly of the fermions is again replaced by the classical anomaly of the WZW model, and the WZW model is now nontrivially fibered over the base GLSM.  The result is a hybrid model of precisely the form discussed in this paper.  Note, too, that this duality has a more familiar name -- it is nothing other than a Narain T-duality of the heterotic string on $X\times (G_{WZW}/H_{R})$ \cite{Evslin:2008zm}.

\section{Some Examples} \label{examples}
We now present some basic examples of WZW models fibered over gauged linear sigma models.  For simplicity, we will take the base space to be a non-compact projective space or Grassmanian.  As usual \cite{Witten:1993yc}, superpotentials can be turned on to cut out a hypersurface/intersection and thus compactify the target.


\subsection{$U(1) \times U(1) \longrightarrow K3$}

Consider a $U(1)$ GLSM for $K3$ decorated by some vector bundle, $\CV\to K3$, such that the gauge anomaly $\CA=c_{2}(T_{K3})-c_{2}(\CV)$ is non-zero.  As we have seen, we can cancel this anomaly by tensoring in a WZW model with suitable left-$U(1)$ action gauged. The simplest such WZW-fiber we can add while preserving \ZT\  \susy\ is the $G=U(1) \times U(1)\sim T^{2}$ WZW model.  To cancel the anomaly, we gauge this WZW model by a left-acting $U(1)$, 
\bea
H_L=U(1)=\open e^{i \alpha N_1} & 0 \\ 0 & e^{i \alpha N_2} \close \, .
\eea
The bosonic Lagrangian then takes the form,
\bea
\mathcal{L}_{fiber}&=&\frac{k}{4 \pi} ( \partial_+ \theta_l \partial_- \theta_l -  2 N_l A_+ \partial_-\theta_l +  (N_1^2 +N_2^2) A_+ A_-) \nonumber \\
&=&\frac{k}{4 \pi} ( D_+ \theta_l D_- \theta_l -  N_l  \theta_l  F_{+-}) \nonumber
\eea
where we have chosen bosonic coordinates $g=(e^{i \theta_1}, e^{i \theta_2}) \in G$ such that $D \theta_l \equiv \partial \theta_l- N_l A$, and we have integrated by parts in the second equality.   The abelian anomaly is canceled by requiring $k(N_1^2+N_2^2)=\CA$.  In terms of the complexified coordinates $\t = \t_{1}+i\t_{2}$ and $\chi={1\over\sqrt{2}}(\psi_{+}^{1}+i\psi_{+}^{2})$, the SUSY transformations of the WZW fields become,
\bea
\d \t &=& \sqrt{2} \,\eps\, \chi \non \\
\d \chi &=& -i\frac{\bar{\eps}}{\sqrt{2}} ( \p_+\t - (N_1+i N_2) A_+) .
\eea
This is nothing but a the torsion linear sigma model of \cite{Adams:2006kb}, a worldsheet description of heterotic flux vacua first explored in  \cite{Dasgupta:1999ss,Fu:2005sm, Fu:2006vj,Becker:2006et}\ whose semi-classical geometry is a \nK\ $T^{2}$-fibration $T^{2}\to X\stackrel{\pi}{\to} K3$ decorated by a vector bundle $\CV_{X}=\pi^{*}\CV_{K3}$ supported by NS-NS 3-form flux $H$ on the total space $X$.  

The realization of these earlier abelian linear models as special cases of WZW-fibrations clarifies a number of features obscured in the earlier presentation.  First, it is now clear why these models arise via bosonization and fermionization -- indeed, that is how the original WZW construction arose.  Secondly, and importantly, the WZW presentation makes precise one of the suggestive features of the original linear models -- namely, the gauge action on the bosonic coordinates on the fiber is chiral, with only the left-action gauged.  This plays an important role in the study of ``small-radius'' phases of the worldsheet theory \cite{Adams2009}.

Before moving on to a non-abelian example, it is perhaps useful to give a concrete example of a model in which all of the anomalies are explicitly cancelled.  We take a particularly simple example -- a $T^2$ fibration over a $K3$ formed by the quartic in $\mathbb{P}^3$. The field content is five chiral multiplets, $\Phi_{i=1, \ldots, 4}$, $P$, and five fermi multiplets, $\Lambda_{a=1, 2,3,4}$ and $\Gamma$, and one WZW multplet $(\t, \psi)$ which forms a $k=1$, $U(1) \times U(1)$, WZW model.   The various charge 
\begin{wrapfigure}[16]{l}{0pt}
\begin{tabular}{|r|r|r|r|}
\hline
Field&Gauge& $U(1)_R$ &$U(1)_L$ \\
\hline
$\Phi_i$& $1$&$0$&$0$\\
$P$ & $-4$&$1$&$1$ \\
$\Lambda_{1,2}$ & $1$&$0$&$-1$\\
$\Lambda_{3,4}$ & $0$ &$0$&$-1$\\
$\Gamma$ &$-4$&$1$&$0$\\
$\theta_1$ &$1$&$0$&$-1$\\
$\theta_2$ &$1$&$0$&$-1$\\
$\psi$ &$0$&$1$&$0$\\
\hline
\end{tabular}
\caption{Charges}
\noindent  ~ \\ 
\label{test}
\end{wrapfigure}
assignments are given in the Figure~1.  
To make the target space compact, we also add a superpotential of the form,
\be
W  \propto \int d \theta^+ \,  (\Gamma \, G(\phi) + P \, \Lambda_a J^a (\phi))  
\ee
where $G(\phi)$ is a quartic polynomial which cuts out a $K3$ in $\mathbb{P}^3$. The $J^a$'s are cubic and quartic polynomials that ensure transversality and set $p=0$ in the $r \gg 1$ phase.

Thus, in the usual way, the $D$ terms and $F$ terms conspire to give a $K3$ over which the $T^2$ is fibered. The model also comes equipped with a gauge bundle, $V$, from the Fermi multiplets, determined by the exact sequence
\begin{equation} 0 \longrightarrow V\longrightarrow \mathcal{O}(1)^5 \xrightarrow{J^a} \mathcal{O}(4)\longrightarrow 0\end{equation}
This was one of the many models studied in  \cite{Adams:2006kb}. 

~

\subsection{Another Abelian Anomaly: $SU(2)\!\times\!U(1) \to \widetilde{\IC^{2}/\IZ_{2}}$}

Consider a $U(1)$ GLSM including two chiral multiplets, $\Phi_i$, of charge $+1$, one chiral multiplet, $P$, of charge $-2$, and one Fermi multiplet, $\Gamma$, of charge $-2$.   The classical higgs branch of this theory is the small resolution of $\IC^{2}/\IZ_{2}$, \ie\ $\CO(-2) \longrightarrow \IP^1$.  Quantum mechanically, this model has a chiral anomaly, so we need to couple in a WZW model.  

Since $\wtilde{\IC^{2}/\IZ_{2}}$ is just a non-compact $K3$, we could again add a $U(1)\times U(1)$ WZW model as in the previous example (this gives a particularly simple and tractable non-compact example).  Instead, let's try fibering over our target a non-abelian WZW model for some group manifold, $G$, with a $U(1)\subset G$ gauged so as to cancel the abelian anomaly of the GLSM.  A particularly simple choice is $G=SU(2)\times U(1)$.  Since 
$SU(2) \times U(1)$ is hyperkahler \cite{Ivanov:1994ec}, the WZW model admits \ZT\ \susy. In particular, the Lie algebra splits as $T_{\pm}$ under three inequivalent complex structures. For example, under one of them
\be
T_\pm= \{ a (1\pm i \sigma_x) + b (\mp \sigma_z + i \sigma_y) \} .
\ee
To cancel the anomaly of  the GLSM, we gauge the left-action of the $U(1)$ factor in WZW model by 
\be
A_L=N  A  
\ee
where $A$ is the vector in the GLSM and $N$ is a parameter.  Anomaly cancellation then fixes $k=2$ and $N=1$.  A simple computation confirms that this model is completely non-anomalous.  The central charge (over three) of WZW model is \cite{Rocek:1991az} $\hat{c}= 2 \frac{k+1}{k+2}=3/2$. The na\"ive central charge of this model is thus $\hat{c}=2 + 3/2=7/2$.  That this na\"ive counting is indeed correct can be seen by flowing to the Landau-Ginsburg point in the moduli space, $r \rightarrow -\infty$. As discussed in \cite{Adams2009}, the correct description of the theory here is an asymmetric orbifold of a $\IC^2$ theory tensored with an $SU(2) \times U(1)$ WZW theory. Since orbifolding by a finite group does not change the central charge \cite{Ginsparg:1988ui}, the central charge is just the sum of the central charges of those two theories. 



\subsection{Examples With Non-Abelian GLSMs}
\begin{wrapfigure}[12]{l}{0pt}
\begin{tabular}{|r|r|}
\hline
Field&Gauge\\
\hline
$\Phi_i$& $\Box$\\
$P_\alpha$ & $-q_\alpha$ \\
$\Lambda^m$ & $\Box$ \\
$\Gamma^{s}$ &$-d_{s}$\\
$\Sigma^{\sigma}$ &$adj.$\\
\hline
\end{tabular}
\caption{\!$U(N_{c})$}
\label{test}
\end{wrapfigure}
Let's now start with a $U(N_c)$ gauge theory of the form studied in \cite{Hori:2006dk}.  These models 
include $N_\Phi$ chiral multiplets $\Phi_i$ transforming in the fundamental, $N_P$ chiral multiplets $P_\alpha$ in the det$^{-q_{\a}}$ representation, $N_\LL$ Fermi multiplets, $\LL^m$ ,  in the fundamental and $N_\G$ Fermi multiplets, $\Gamma^s$,  in the det$^{-d_{s}}$ representation. In addition, we add $N_\Sigma$ chiral multiplets, $\Sigma_\sigma$, in the adjoint representation.  The field content is summarized in Figure~2.

The classical target space is given by the vanishing locus of the $D$ term,
\be\label{Grass}  
D^a_b= e^2 \left( \sum_{i=1}^{N_R} \phi_i^a \phi^\dagger_{b i} -\sum_\a q_\a |p_\a|^2 \delta^a_b - r \delta^a_b \right) \, \, \, a,b=1 \ldots N_c \, ,
\ee
modulo the gauge group, as usual.  On the Higgs branch, where $p^\a=0$, the manifold defined by $D=0$ is the space of $N_c$ planes in $\mathbb{C}^{N_R}$, also known as the Grassmannian $G(N_c, N_R)$. In the non-anomalous models of  \cite{Hori:2006dk} , a superpotential restricts the vacuum manifold to be some Calabi-Yau hypersurface of $G(N_c, N_R)$.  For our purposes, the non-compact ambient variety suffices, so we will dispense with the superpotential.   

To study the anomaly structure of the theory, it is useful to treat the trace and traceless parts of the gauge group separately.  For gauge transformations in the $SU(N)$, 
\be \label{nonabeliananomaly} 
\partial_\mu j^\mu= \frac{N_R-N_L-2 N_c+2N_\sigma N_c}{4 \pi}\, Tr(\alpha F'_{+-})  \, .
\ee
Note that for $SU(N_c)$, $Tr_{Adjoint} (T^a T^b)=2N_c Tr(T^a T^b)$.
For gauge transformations in the central $U(1)$, on the other hand, 
\be
\partial_\mu j^\mu= \frac{1}{4 \pi} (N_R-N_L+ N_c \sum_\alpha q_\alpha^2 -N_c \sum_s d_s^2 ) \, Tr(\alpha F'_{+-}) .
\ee
To cancel these anomalies, we again tensor in and gauge a suitable WZW model.  If the non-abelian anomaly is non-trivial, however, the WZW model must also be non-abelian.  Let's look at a couple of simple examples.

\subsubsection{$SU(2)\!\times\!U(1) \to \[\oplus_\alpha O(-q_\alpha) \to G(2, N_R)\]$}

As in a previous example, we start with a $SU(2) \times U(1)$ WZW model, but this time gauge the entire symmetry group (which we identify with the gauge group of the non-abelian GLSM),
\be
A_L=N_0 T^0 A^0 +  T^a A^a 
\ee
where $a=1,2,3$ runs over the $SU(2)$ generators and $0$ denotes the central $U(1)$ in both the WZW model and the GLSM. The anomaly is cancelled by requiring 
\bea
k N_0^2  &=&   N_R   -N_L  + 2 \sum_\a q_\a^2   -2 \sum_s d_s^2    \non \\
k   &=&   N_R  - N_L  -4 +4N_\sigma .
\eea


\subsubsection{($[U(1)^2]_{k'} \times [SU(2)/U(1)]_k) \to [\oplus_\a O(-q_\a) \to G(2, N_R)]$}
We present a second way of fibering a WZW model over a non-abelian gauge theory--one that utilizes both left and right gauging of the WZW model. Starting with a GLSM of the same form as in the previous example, we now cancel the anomaly by tensoring in and left-gauging an $[SU(2)/U(1)]\!\otimes\![U(1)^{2}]$ WZW model.

As above, a $U(1)$ subgroup of the WZW model is left-gauged by the central $U(1)$ of the GLSM, with two integers, $\{N_1, N_2\}$, specifying the embedding of $U(1)$ in $U(1)^2$ s.t. the abelian anomaly is cancelled.  The full $SU(2)$ of the second WZW model is also left-gauged by the $SU(2)$ vector of the GLSM, canceling the non-abelian anomaly.  Finally, to cancel the anomaly of $H_{R} \sim U(1)$, we also add a left-moving fermion with charged $Q$ under the auxiliary $U(1)_{right}$ gauge symmetry of the WZW model. The full anomaly cancellation conditions are thus,
\bea
k' (N_1^2+N_2^2)    &=&    N_R - N_L  + 2 \sum_\a q_\a^2  -2 \sum_s d_s^2 \non \\
k &=& N_R -N_L -4 +4N_\s \non \\
k &=& 2 (Q^2-1). 
\eea




\begin{comment}
\subsection{$U(1) \times U(1) \longrightarrow \mathbb{P}^1$}
We are still very confused about this guy! ...section not fit for reading \\ \\

There is no restriction to CY base, as we will now demonstrate. We take a toy $k=2$ $U(1) \times U(1)$ WZW model over $\mathbb{P}^1$. The matter content is two chiral multiplets $\Phi_i$  and a torsion multiplet  $\theta_{1,2}$, $\psi$. The charge assignments are
\\ \\
\begin{tabular}{|r|r|r|r|}
\hline
Field&Gauge&$U(1)_R$&$U(1)_L$\\
\hline
$\Phi_i$& $+1$&$r$&$r$\\
$\theta_1$ & $1$&$q_{1R}$&$q_{1L}$ \\
$\theta_2$ &$0$&$q_{2R}$&$q_{2L}$\\
$\psi_{R\pm}$ &$0$&$1$&$0$\\
\hline
\end{tabular} \\ \\
The gauge anomaly cancels, as $2*1^2=k(N_1^2 +N_2^2)=2$. The conditions for $U(1)_R$/anomaly, $U(1)_L$/anomaly cancellations are, respectively

\begin{eqnarray} (r-1)-2 q_{1R}&=&0 \nonumber \\
r-2 q_{1L}&=&0 \end{eqnarray}
Again, I am stifled by the global anomaly cancellation.

Reasons why we expect this to be $SU(2) \times U(1)$ WZW model....
\end{comment}

\section{Conclusions}

In this note we have shown that anomalies in \ZT\ gauged linear sigma models may be cancelled by tensoring them with a suitably gauged WZW model.  The resulting gauged WZW+LSM is manifestly $\CN=2$ \susic\ and is expected to flow to a non-linear sigma model with NS-NS flux when the mixed gauge-$R$-anomaly is also vanishing.  Along the way we identified a candidate $R$-current which is in the same $Q$-cohomology class as the $R$-current of the twisted SCFT, and is thus expected to flow to the superconformal $R$-current of the IR SCFT.  We also found that these WZW models reduce, in the abelian case, to the ``torsion linear sigma models'' of \cite{Adams:2006kb}; the more general non-abelian case thus provides a natural generalization of these quasi-geometric heterotic flux vacua.

It is straightforward (if tedious) to integrate out the massive vector and matter fields along the semi-classical Higgs branch to construct a one-loop approximation to the geometry and flux of the sigma model to which the gauge theory flows. (We must work at one-loop rather than tree level due to the anomaly.)  As in the abelian case studied in detail in \cite{Adams:2006kb}, the result is again a \nK\ metric with flux specified by the WZW-fibration and satisfying the Bianchi identity.  Moreover, one should be able to explicitly identify the cohomology classes on the base specifying the full flux vacuum by studying various twisted models and their chiral rings \cite{Adams:2005tc,Melnikov:2009nh}.  It would be interesting to study this quasi-geometry in detail.  


\vspace{1cm}
{\bf Acknowledgements}
We thank 
Jacques Distler, 
Shamit Kachru, 
Josh Lapan,
Albion Lawrence,
John McGreevy
and
Eric Sharpe
for illuminating discussions.
A.A. thanks the organizers and participants of the 2008 Aspen Center for Physics where some of this work was discussed.
This work was supported in part by funds provided by the U.S. Department of Energy
(D.O.E.) under cooperative research agreement DE-FG0205ER41360.

\renewcommand{\theequation}{\Alph{section}.\arabic{equation}}

\setcounter{equation}{0}

\bibliographystyle{utphys}
\bibliography{WZWnotesbib}

\end{document}